# NUMERICAL CASEOLOGY BY LAGRANGE INTERPOLATION FOR THE 1D NEUTRON TRANSPORT EQUATION IN A SLAB

B.D. Ganapol[1]

*University of Arizona*
*Department of Aerospace and Mechanical Engineering*
*Tucson AZ 85712, USA*

Ganapol@cowboy.ame.arizona.edu

## ABSTRACT

Here, we are concerned with a new, highly precise, numerical solution to the 1D neutron transport equation based on Case's analytical singular eigenfunction expansion (SEE). While a considerable number numerical solutions currently exist, understandably, because of its complexity, even in 1D, there are only a few truly analytical solutions to the neutron transport equation. In 1960, Case introduced a consistent theory of the SEE for a variety of idealized transport problems and forever changed the landscape of analytical transport theory. Several numerical methods including the F$_N$ method were based on the theory. What is presented is yet another, called the Lagrange order *N* method (LNM), featuring the simplicity and precision of the F$_N$ method, but for a more convenient and natural Lagrangian polynomial basis.

## I. INTRODUCTION

Because of full-range orthogonality, Case's method [1], commonly known as Caseology, produces an elegant analytical form of solution to the monoenergetic 1D neutron transport equation by following the Fourier approach for solving partial differential equations. In particular, a complete set of basis functions to the transport equation through separation of variables define the eigenmodes of the homogeneous equation. The general solution is then expanded in the complete set and their orthogonality gives the expansion coefficients. The curiosity of the expansion is that the operator spectrum consist of a finite number of discrete and a continuum set of eigenmodes.

Many individuals contributed to the development of singular eigenfunction solutions to various transport equations. Davison [2], commonly acknowledged as the originator of the approach, worked out some of the early proofs of concept in the mid 40's. Other contributors include Lafore and Millot [3], Wigner [4], Van Kampen [5] who was the first to apply the theory to plasmas. But it was Case's 1960 paper [1] that brought the theory together in a consistent fashion to analytically solve the most fundamental problems in transport theory. The method's popularity peaked in the '60's and 70's with the appearance of Case and Zweifel's classic book, *Linear Transport*

*Theory* [6], which was essentially the second volume of Case de Hoffman and Placzek's *Introduction to Neutron Diffusion* [7]. During that time, numerous papers appeared on Caseology in many fields of physics including, of course, nuclear reactor theory, acoustics, radiative transfer, rarefied gas dynamics, kinetic theory, traffic flow, and graphics to name several.

The value of Caseology, is mainly theoretical to enable the solution of the transport equation to be as close to a closed form exact solution as possible. For monoenergetic neutron transport with isotropic scattering, exclusively considered here, there are only a handful of problems that lend themselves to closed forms, *i.e.*, primarily transport in an infinite medium and a half-space. A second advantage of Caseology, is the foundation it provides to develop highly precise numerical solutions, such as the F$_N$ method. To the author's knowledge, the numerical solution, to be presented, is different from those found in the literature. When expressed as the full-range SEE, the solution for the slab is no longer in a closed form unlike for infinite media. Its evaluation involves iteration through auxiliary X- and Y- functions as in invariant imbedding or Fredholm integral equations to solve non-linear coupled integral equations. The same is true for the half-space to determine the Case *X*-function (or equivalently Chandrasekhar's *H*-function), but unlike the slab, these functions have closed form integral representations. Our goal is to avoid evaluation of the auxiliary non-linear integral equations for the slab and remain true to the original form of the full-range SEE. While SEE is not the only way to find analytical solutions, in comparison to Weiner-Hopf, invariant imbedding, Laplace and Fourier transforms, arguably it is the most mathematically pleasing.

A full-range approach coupled to Lagrange interpolation will determine the numerical solution for the slab. One may find the approach algebraically intimidating, but the motivation is clear. Full-range not only avoids the auxiliary non-linear integral equations; but also avoids half-range orthogonality, requiring knowledge of solutions to singular integral equations (SIEs). The identical approach was taken by Siewert [8], who cleverly devised the F$_N$ method, based on a full-range scheme, to construct a spectral expansion through collocation. A subsequent version, bypassed Case's method altogether by expressing the transport equation for the exiting flux directly as SIEs.

Since the F$_N$ method is one of the most successful methods of solving the 1D neutron and radiative transfer equations on the planet, one might ask-- Why should an additional numerical solution be of interest? First, the LNM expresses SEE in a new form by decoupling the solution of the SIEs for the continuum coefficients from the discrete. Second, the required Lagrange polynomial basis is a natural choice that accommodates a half-range numerical formulation and seamlessly couples to Gauss Quadrature (GQ).

The LNM gives excellent agreement with other methods such as the Response Matrix/Discrete Ordinates Method (RM/DOM), Adding and Doubling Method (ADM), the Double PN method (DPN) and Matrix Riccati Equation Method (MREM) published by the author [9,10,11,12]. Best results however, seem to be for continuous source distributions, in particular an isotropic source in predominantly scattering media, though acceptable benchmark precision is also found for the perpendicular beam.

A word on the perspective of what is about to be presented. Unlike recent efforts of the author (RM/DOM, DPN, ADM, MREM) to solve the 1D transport equation with anisotropic scattering, the LNM is limited to isotropic scattering for several reasons. First, obviously, it is always easier to demonstrate a new numerical method for isotropic scattering. The second reason is one of simplicity in calculating the discrete eigenvalues. The determination of the discrete eigenvalues for anisotropic scattering kernels, whose number is generally unknown, is difficult because of the potential variability of scattering kernels. This becomes particularly challenging when absorption is small as the eigenvalues shed from the continuum spectrum. So we consider only isotropic scattering to emphasize the new formulation. At some later time however, the theory for anisotropic scattering will be attempted.

We begin by representing Case's solution in terms of the discrete and singular eigenfunctions for an isotropically scattering slab. The known incoming flux entering the slab surfaces provide two coupled singular equations for the expansion coefficients. By adding and subtracting, the equations uncouple to give a combination of just the continuum coefficients in terms of the discrete. In addition, superposition eliminates the dependence on the discrete coefficients enabling the continuum coefficients to be determined independently from the discrete. The discrete coefficients come from orthogonality requiring integration over the continuum coefficients already found by combining Lagrange interpolation with GQ.

## II. THEORY

### II.A. Singular Integral Equations for Continuum Expansion Coefficients

Our focus is the solution to the following transport equation for a slab of width $\Delta$, measured in mean free paths (*mfp*):

$$\left[\mu\frac{\partial}{\partial x}+1\right]\psi(x,\mu)=\frac{c}{2}\int_{-1}^{1}d\mu'\psi(x,\mu'), \tag{1a}$$

with entering flux $\psi(\mu)$ (the source) at near surface ($x = 0$) and no entering flux at the far surface ($x = \Delta$)

$$\begin{aligned}\psi(0,\mu) &\equiv \psi(\mu) \\ \psi(\Delta,-\mu) &\equiv 0\end{aligned} \tag{1b,c}$$

for $0 \leq \mu \leq 1$. Case's method gives the full-range SEE solution [1] for the homogeneous transport equation as

$$\psi(x,\mu)=a_{0+}\phi_{0+}(\mu)e^{-x/v_0}+a_{0-}\phi_{0-}(\mu)e^{x/v_0}+\int_{-1}^{1}dve^{-x/v}\phi_v(\mu)A(v), \tag{2a}$$

with singular eigenfunctions (SEs)

$$\phi_{0+}(\mu) = \frac{cv_0}{2} \frac{1}{v_0 - \mu}$$

$$\phi_v(\mu) = \frac{cv}{2} P \frac{1}{v - \mu} + \lambda(v)\delta(v - \mu),$$

(2b,c)

where $P$ indicates the principal value under an integral, and

$$\lambda(v) = 1 - \frac{cv}{2} \log\left[\left|\frac{v+1}{v-1}\right|\right]. \tag{2d}$$

The SEs satisfy symmetry relations

$$\phi_{0\pm}(\mu) = \phi_{0+}(\pm\mu)$$
$$\phi_{v\pm}(\mu) = \phi_v(\pm\mu),$$

(3a,b)$^{\pm}$

where $v_0$ is the discrete eigenvalue that satisfies the dispersion relation

$$\lambda(v_0) = 1 - \frac{cv_0}{2} \log\left[\frac{v_0+1}{v_0-1}\right] = 0. \tag{3c}$$

The expansion coefficients, $a_{0+}$, $a_{0-}$, $A(v)$, can be explicitly determined from the orthogonality of the SEs

$$\int_{-1}^{1} d\mu\mu\phi_v(\mu)\phi_{v'}(\mu) = N_v \delta(v - v')$$

$$N_{v\pm} \equiv \pm v\left[\lambda^2(v) + \left(\frac{c\pi v}{2}\right)^2\right]$$

(4a,b$^{\pm}$,c$^{\pm}$)

$$\int_{-1}^{1} d\mu\mu\phi_{0\pm}(\mu)^2 = \pm N_{0+} \equiv \pm\left[\frac{cv_0}{2}\left[\frac{c}{1 - 1/v_0^2}\right] - 1\right].$$

but will be found by an alternative procedure.

Introducing the BCs at the two free surfaces for $0 \le \mu \le 1$ into Eq(2a) for $x = 0$, gives

$$\psi(\mu) = a_{0+}\phi_{0+}(\mu) + a_{0-}\phi_{0+}(-\mu) + \int_0^1 dv\, \phi_v(\mu) A^+(v) + \int_0^1 dv\, \phi_v(-\mu) A^-(v), \quad (5a)$$

and for $x = \Delta$,

$$0 = a_{0+}e_{0-}\phi_{0+}(-\mu) + a_{0-}e_{0+}\phi_{0+}(\mu) + \int_0^1 dv \left[ \begin{array}{l} e_{v-}\phi_v(-\mu) A^+(v) + \\ +e_{v+}\phi_v(\mu) A^-(v) \end{array} \right] \quad (5b)$$

with

$$\begin{aligned} A^{\pm}(v) &\equiv A(\pm v) \\ e_{0\pm} &\equiv e^{\pm \Delta/v_0} \\ e_{v\pm} &\equiv e^{\pm \Delta/v}. \end{aligned} \quad (5c,d,e)^{\pm}$$

By adding and subtracting Eqs(5a,b), there results, after changing $\mu$ to $v$

$$\psi(v) = \left[\phi_{0+}(v) \pm e_{0-}\phi_{0+}(-v)\right] b_0^{\pm} + \int_0^1 dv' \left[\phi_{v'}(v) \pm e_{v'-}\phi_{v'}(-v)\right] B^{\pm}(v'), \quad (6a)^{\pm}$$

where the discrete and continuum coefficients combine as

$$\begin{aligned} b_0^{\pm} &\equiv a_{0+} \pm e_{0+}a_{0-} \\ B^{\pm}(v) &\equiv A^+(v) \pm e_{v+}A^-(v), \end{aligned} \quad (6b,c)^{\pm}$$

and after transposition of the first term, the following two uncoupled SIEs result for $0 \leq v \leq 1$:

$$\lambda(v) B^{\pm}(v) + \frac{c}{2} \int_0^1 dv' \frac{v'}{v'-v} B^{\pm}(v') \pm \frac{c}{2} \int_0^1 dv' \frac{v'}{v'+v} e_{v'-} B^{\pm}(v') =$$
$$= \psi(v) - u_0^{\pm}(v) b_0^{\pm}, \quad (7a)^{\pm}$$

with

$$u_0^{\pm}(v) \equiv \phi_{0+}(v) \pm e_{0-}\phi_{0+}(-v). \quad (7b)^{\pm}$$

The horizontal line in the integral symbol indicates a principal value integration.

Using linearity, we can further split the SIEs into

$$\lambda(v)B_i^\pm(v) + \frac{c}{2}\int_0^1 dv' \frac{v'}{v'-v} B_i^\pm(v') \pm \frac{c}{2}\int_0^1 dv' \frac{v'}{v'+v} e_{v'_-} B_i^\pm(v') =$$

$$= f_i^\pm(v) \equiv \begin{cases} \psi(v), & i=1 \\ u_0^\pm(v), & i=2 \end{cases}$$

(8a)$^\pm$

and re-sum to give

$$B^\pm(v) = B_1^\pm(v) - b_0^\pm B_2^\pm(v).$$

(8b)$^\pm$

Note the linear dependence separates the unknown discrete coefficient $b_0^\pm$ from the unknown continuum coefficients in $B_i^\pm(v)$. Eqs(8a)$^\pm$ are a key feature of our formulation since they eliminate the need to know $b_0^\pm$ to find $B_i^\pm(v)$, the combination of continuum coefficients. Now, on to solve Eqs(8a)$^\pm$.

## II.B. Lagrange Interpolation for $B_i^\pm(v)$

From Lagrange interpolation at $N$ nodes $v_j$

$$B_i^\pm(v) = \sum_{j=1}^N B_{ij}^\pm l_j(v) = \sum_{j=1}^N B_{ij}^\pm \left[\frac{P_N^*(v)}{P_N^{*\prime}(v_j)(v-v_j)}\right].$$

(9a)$^\pm$

$v_j$, $j=1,...,N$ are the zeros of the shifted Legendre polynomial $P_N^*(v_j)$

$$P_N^*(v_j) \equiv P_N(2v_j - 1) = 0;$$

(9b)

and its derivative is

$$P_N^{*\prime}(v_j) \equiv 2P_N'(z)_{z=2v_j-1},$$

(9c)

where $P_N(2v-1)$ is the standard Legendre polynomial. Substitution of Eq(9a)$^\pm$ into the first integral of Eqs(8a)$^\pm$ and using Gauss quadrature for the second, since it is not a principal value integral, gives

$$\lambda(v)B_i^\pm(v) + \frac{c}{2}\sum_{j=1}^{N}\left[\frac{I(v,v_j)}{P_N^{*\prime}(v_j)} \pm \omega_j e_{v_j-}\frac{v_j}{v_j+v}\right]B_{ij}^\pm = f_i^\pm(v), \quad \text{(10a,b)}^\pm$$

where $\omega_j$ are the Gauss Quadrature weights and

$$I(v,v_j) \equiv \int_0^1 dv' \frac{v'}{v'-v}l_j(v') = -2\begin{cases}\frac{1}{v_j-v}\left[v_j Q_N(2v_j-1) - vQ_N(2v-1)\right], & v \neq v_j \\ Q_N(2v_j-1) + 2v_j Q_N'(z)\big|_{z=2v_j-1}, & v = v_j.\end{cases} \quad (10c)$$

$Q_N(v)$ is the Legendre function of the second kind of order $N$. Note that Lagrange interpolation combined with Gauss quadrature is quite convenient if the interpolation abscissae are identical to the quadrature abscissae as is true here. This is the second key feature of LNM. In addition, the advantage of the interpolation is that the principal value integration in the first integral is treated exactly giving rise to the $Q$-functions in $I(v,v_j)$.

Finally, letting $v = v_m$, $m = 1,...,N$ leads to the following set of linear equations, which when solved give $B_{im}^\pm(v)$, $m = 1,...,N$, $i = 1,2$

$$\sum_{j=1}^{N}\left\{\left[\lambda(v_m) - \frac{c}{2}\frac{I(v_m,v_m)}{P_N^{*\prime}(v)}\right]\delta_{jm} + \frac{c}{2}\left[(1-\delta_{jm})\frac{I(v_m,v_j)}{P_N^{*\prime}(v)} \pm \omega_j\frac{v_j}{v_j+v_m}e_{v_j-}\right]\right\}B_{ij}^\pm = f_i^\pm(v_m).$$

$$(11)^\pm$$

Thus, from Eq(6c)$^\pm$, the continuum coefficients, $A^\pm(v_m)$, can be found. As will be shown, this will be unnecessary.

### II.C. Determination of Discrete Coefficients $b_0^\pm$

From the analytical expression for the flux given by Eq(2a)

$$\psi(x,\mu) = a_{0+}\phi_{0+}(\mu)e^{-x/v_0} + a_{0-}\phi_{0-}(\mu)e^{x/v_0} + \int_{-1}^{1}dv\, e^{-x/v}\phi_v(\mu)A(v) \quad (12)$$

with $\mu$ replaced by $-\mu$ at the near surface, $x = 0$, and with $x = \Delta$ at the far surface, one finds

$$\psi(0,-\mu) = a_{0+}\phi_{0+}(-\mu) + a_{0-}\phi_{0+}(\mu) +$$

$$+ \int_0^1 dv \left[\phi_\nu(-\mu) A^+(\nu) + \phi_\nu(\mu) A^-(\nu)\right] \quad (13a)$$

$$\psi(\Delta,\mu) = a_{0+}\phi_{0+}(\mu)e_{0-} + a_{0-}\phi_{0+}(-\mu)e_{0+} +$$

$$+ \int_0^1 dv \left[e_{\nu-}\phi_\nu(\mu) A^+(\nu) + e_{\nu+}\phi_\nu(-\mu) A^-(\nu)\right] \quad (13b)$$

over the full-range $-1 \le \mu \le 1$. Then adding and subtracting

$$\psi(0,-\mu) \pm \psi(\Delta,\mu) = \left[\phi_{0+}(-\mu) \pm e_{0-}\phi_{0+}(\mu)\right] b_0^\pm +$$

$$+ \int_0^1 dv \left[\phi_\nu(-\mu) \pm e_{\nu-}\phi_\nu(\mu)\right] B^\pm(\nu) \quad (14a)^\pm$$

and from orthogonality by projection over $\mu\phi_{0+}(-\mu)$

$$-N_{0+} b_0^\pm = \int_{-1}^{1} d\mu\mu\phi_{0+}(-\mu)\left[\psi(0,-\mu) \pm \psi(\Delta,\mu)\right] \quad (14b)^\pm$$

using Eqs(4). Next, separating into half-ranges

$$-N_{0+} b_0^\pm = \int_0^1 d\mu\mu\phi_{0+}(-\mu)\left[\psi(0,-\mu) \pm \psi(\Delta,\mu)\right] - \int_0^1 d\mu\mu\phi_{0+}(\mu)\psi(\mu), \quad (15)^\pm$$

where we have introduced the BC [Eqs(1b,c)] into the second integral. When Eqs(14a)$^\pm$ are introduced into the first integral

$$-N_{0+} b_0^\pm = -J_0^+ +$$

$$+ \int_0^1 d\mu\mu\phi_{0+}(-\mu) \begin{bmatrix} \left[\phi_{0+}(-\mu) \pm e_{0-}\phi_{0+}(\mu)\right] b_0^\pm + \\ + \int_0^1 dv \left[\phi_\nu(-\mu) \pm e_{\nu-}\phi_\nu(\mu)\right] B^\pm(\nu) \end{bmatrix} \quad (16a)^\pm$$

or more compactly

$$N_{0+}b_0^\pm = J_0^+ - T_1^\pm b_0^\pm - \int_0^1 d\nu T_2^\pm(\nu)\left[B_1^\pm(\nu) - b_0^\pm B_2^\pm(\nu)\right] \quad (16b)^\pm$$

with

$$J_0^+ \equiv \int_0^1 d\mu\mu\phi_{0+}(\mu)\psi(\mu)$$

$$T_1^\pm \equiv \int_0^1 d\mu\mu\phi_{0+}(-\mu)\left[\phi_{0+}(-\mu) \pm e_{0-}\phi_{0+}(\mu)\right] \quad (16c,d^\pm,e^\pm)$$

$$T_2^\pm(\nu) \equiv \int_0^1 d\mu\mu\phi_{0+}(-\mu)\left[\phi_\nu(-\mu) \pm e_{\nu-}\phi_\nu(\mu)\right]$$

and substituting Eq(8b)$^\pm$ for $B^\pm(\nu)$. Five half-range integrals are required in Eqs(16c,d$^\pm$,e$^\pm$), all of which can be done analytically once $\psi(\mu)$ is specified. Thus, for $T_1^\pm$

$$T_1^\pm = T_{11} \pm e_{0-}T_{12}$$

$$T_{11} \equiv \int_0^1 d\mu\mu\phi_{0+}(-\mu)\phi_{0+}(-\mu) = \left(\frac{c\nu_0}{2}\right)^2\left[\ln\frac{\nu_0+1}{\nu_0} - \frac{1}{\nu_0+1}\right] \quad (17a^\pm,b,c)$$

$$T_{12} \equiv \int_0^1 d\mu\mu\phi_{0+}(-\mu)\phi_{0+}(\mu) = \frac{1}{2}\frac{c\nu_0}{2}\left[1 - c\nu_0 \ln\frac{\nu_0+1}{\nu_0}\right].$$

Similarly, for $T_2^\pm(\nu)$

$$T_2^\pm(\nu) = T_{21}(\nu) \pm e_{\nu-}T_{22}(\nu)$$

$$T_{21}(\nu) \equiv \int_0^1 d\mu\mu\phi_{0+}(-\mu)\phi_\nu(-\mu) = \phi_{0+}(\nu)\left(\frac{c\nu_0}{2}\right)\left[\nu_0\ln\frac{\nu_0+1}{\nu_0} - \nu\ln\frac{\nu+1}{\nu}\right] \quad (18a^\pm,b,c)$$

$$T_{22}(\nu) \equiv \int_0^1 d\mu\mu\phi_{0+}(-\mu)\phi_\nu(\mu) = -\nu\phi_{0+}(-\nu)\left[\left(\frac{c\nu_0}{2}\right)\ln\frac{\nu_0+1}{\nu_0} + \left(\frac{c\nu}{2}\right)\ln\frac{\nu+1}{\nu}\right] +$$

$$+ \nu\phi_{0+}(-\nu)\lambda(\nu)$$

$$= \nu\phi_{0+}(-\nu)\left[1 - \left(\frac{c\nu_0}{2}\right)\ln\frac{\nu_0+1}{\nu_0} + \left(\frac{c\nu}{2}\right)\ln\frac{\nu+1}{\nu}\right].$$

Finally, solving for $b_0^\pm$ in Eqs(16b)$^\pm$ gives

$$b_0^\pm = \left[ N_{0+} + T_1^\pm - \sum_{j=1}^{N} \omega_j T_{2j}^\pm B_{2j} \right]^{-1} \left[ J_0^+ - \sum_{j=1}^{N} \omega_j T_{2j}^\pm B_{1j} \right] \quad (19)^\pm$$

with $B_{ij}^\pm$, $j=1,...,N$, $i=1,2$ known. We have found all the necessary coefficients to evaluate Eq(2a), but first we must arrange Eq(2a) appropriately.

For completeness, projecting Eqs(14a)$^\pm$ over $\mu\phi_{0-}(\mu)$, gives the alternative expression for $b_0^\pm$

$$b_0^\pm = \left[ \pm e_{0-} N_{0+} - T_3^\pm + \int_0^1 dv T_4^\pm(v) B_2^\pm(v) \right]^{-1} \left[ \int_0^1 dv T_4^\pm(v) B_1^\pm(v) - J_0^- \right], \quad (20a)^\pm$$

where

$$J_0^- \equiv -\int_0^1 d\mu\mu\phi_{0+}(-\mu)\psi(\mu)$$

$$T_3^\pm \equiv T_{12} \pm e_{0-} T_{31}$$

$$T_{12} = \int_0^1 d\mu\mu\phi_{0+}(\mu)\phi_{0+}(-\mu) \quad (20\text{b},\text{c}^\pm,\text{d},\text{e})$$

$$T_{31} = \int_0^1 d\mu\mu\phi_{0+}(\mu)\phi_{0+}(\mu)$$

and

$$T_4^\pm(v) \equiv T_{32}(v) \pm e_{v-} T_{33}(v)$$

$$T_{32}(v) = \int_0^1 d\mu\mu\phi_{0+}(\mu)\phi_v(-\mu) \quad (20\text{f}^\pm,\text{g},\text{h})$$

$$T_{33}(v) = \int_0^1 d\mu\mu\phi_{0+}(\mu)\phi_v(\mu).$$

**II.D. Final Expressions for the Exiting Flux**
From Eqs(14a)$^\pm$

$$\psi(0,-\mu) \pm \psi(\Delta,\mu) = \tau_0^{\pm}(\mu) b_0^{\pm} + \int_0^1 dv \tau^{\pm}(v,\mu) B^{\pm}(v), \qquad (21a)^{\pm}$$

with

$$\begin{aligned} \tau_0^{\pm}(\mu) &\equiv \phi_{0-}(\mu) \pm e_{0-}\phi_{0+}(\mu) \\ \tau^{\pm}(v,\mu) &\equiv \phi_v(-\mu) \pm e_{v-}\phi_v(\mu). \end{aligned} \qquad (21b,c)^{\pm}$$

When we introduce the expressions for $\phi_v(\pm\mu)$ into Eqs(21c,d)$^{\pm}$ and subsequently into Eq(21a)$^{\pm}$ for $0 \le \mu \le 1$, there results

$$\psi(0,-\mu) \pm \psi(\Delta,\mu) = \tau_0^{\pm}(\mu) b_0^{\pm} + I_1^{\pm}(\mu) \pm I_2^{\pm}(\mu) \pm e_{\mu-}\lambda(\mu) B^{\pm}(\mu), \qquad (22a)^{\pm}$$

where the integrals are

$$\begin{aligned} I_1^{\pm}(\mu) &\equiv \frac{c}{2}\int_0^1 dv \frac{v}{v+\mu} B^{\pm}(v) \\ I_2^{\pm}(\mu) &\equiv \frac{c}{2}\int_0^1 dv \frac{v}{v-\mu} e_{v-} B^{\pm}(v). \end{aligned} \qquad (22b,c)^{\pm}$$

For $I_2^{\pm}(\mu)$, we can add and subtract to remove the singularity to write

$$I_2^{\pm}(\mu) = \frac{c}{2}\int_0^1 dv v \left[\frac{e_{v-} - e_{\mu-}}{v-\mu}\right] B^{\pm}(v) + e_{\mu-}\frac{c}{2}\int_0^1 dv \frac{v}{v-\mu} B^{\pm}(v). \qquad (23a)^{\pm}$$

Conveniently, from Eqs(7a)$^{\pm}$, the principal value integral in Eq(23a)$^{\pm}$ is

$$\frac{c}{2}\int_0^1 dv \frac{v}{v-\mu} B^{\pm}(v) = -\pm\frac{c}{2}\int_0^1 dv \frac{v}{v+\mu} e_{v-} B^{\pm}(v) + \\ + \psi(\mu) - \lambda(\mu) B^{\pm}(\mu) - u_0^{\pm}(\mu) b_0^{\pm}. \qquad (23b)^{\pm}$$

When introduced into Eq(23a)$^{\pm}$, we find with some algebra

$$\psi(0,-\mu) \pm \psi(\Delta,\mu) = \pm \psi(\mu)e_{\mu-} + \left[\tau_0^{\pm}(\mu) - \pm u_0^{\pm} e_{\mu-}\right]b_0^{\pm} +$$

$$+\frac{c}{2}\int_0^1 dv \frac{v}{v+\mu}\left[1-e_{\mu-}e_{v-}\right]B^{\pm}(v) \pm \frac{c}{2}\int_0^1 dvv\left[\frac{e_{v-}-e_{\mu-}}{v-\mu}\right]B^{\pm}(v). \quad (24)^{\pm}$$

Finally, simply adding and subtracting Eqs(24)$^{\pm}$ and dividing by 2 gives the individual exiting angular flux distributions $\psi(0,-\mu)$ and $\psi(\Delta,\mu)$ for $0 \le \mu \le 1$

$$\psi(0,-\mu) = \frac{1}{2}\left\{\begin{array}{l}\left[\tau_0^+(\mu)-u_0^+ e_{\mu-}\right]b_0^+ + \left[\tau_0^-(\mu)+u_0^- e_{\mu-}\right]b_0^- + \\ +\frac{c}{2}\int_0^1 dv \frac{v}{v+\mu}\left[1-e_{\mu-}e_{v-}\right]\left[B^+(v)+B^-(v)\right] + \\ +\frac{c}{2}\int_0^1 dvv\left[\frac{e_{v-}-e_{\mu-}}{v-\mu}\right]\left[B^+(v)-B^-(v)\right]\end{array}\right\} \quad (25a)$$

$$\psi(\Delta,\mu) = \psi(\mu)e_{\mu-} +$$

$$+\frac{1}{2}\left\{\begin{array}{l}\left[\tau_0^+(\mu)-u_0^+ e_{\mu-}\right]b_0^+ - \left[\tau_0^-(\mu)+u_0^- e_{\mu-}\right]b_0^- + \\ +\frac{c}{2}\int_0^1 dv \frac{v}{v+\mu}\left[1-e_{\mu-}e_{v-}\right]\left[B^+(v)-B^-(v)\right] + \\ +\frac{c}{2}\int_0^1 dvv\left[\frac{e_{v-}-e_{\mu-}}{v-\mu}\right]\left[B^+(v)+B^-(v)\right]\end{array}\right\}, \quad (25b)$$

where all singularities have been resolved.

Note that the first term on the RHS of Eq(25b) is the uncollided flux leaving the far surface.

**II.E. Expressions for Interior Flux**

For completeness the coefficients $a_{0+}$, $a_{0-}$, $A^{\pm}(v)$ come from Eq(6b,c)$^{\pm}$ as

$$a_{0+} = \frac{1}{2}\left[b_0^+ + b_0^-\right], \quad a_{0-} = \frac{e_{0-}}{2}\left[b_0^+ - b_0^-\right]$$

$$A^+(v) = \frac{1}{2}\left[B^+(v)+B^-(v)\right], \quad A^-(v) = \frac{e_{v-}}{2}\left[B^+(v)-B^-(v)\right]; \quad (26a,b)^{\pm}$$

however, their explicit form is not necessary as will now be shown.

One can determine the flux interior in the slab from Eq(2a). First, let $\mu$ be negative and apply Eqs(3)

$$\psi(x,-\mu) = a_{0+}\phi_{0+}(-\mu)e^{-x/v_0} + a_{0-}\phi_{0+}(\mu)e^{x/v_0} +$$
$$+ \int_0^1 dve^{-x/v}\phi_v(-\mu)A^+(v) + \int_0^1 dve^{x/v}\phi_v(\mu)A^-(v), \quad (27a)$$

then let $x$ be $\Delta-x$

$$\psi(\Delta-x,\mu) = a_{0+}e_{0-}\phi_{0+}(\mu)e^{x/v_0} + a_{0-}e_{0+}\phi_{0+}(-\mu)e^{-x/v_0} +$$
$$+ \int_0^1 dve^{x/v}e_{v-}\phi_v(\mu)A^+(v) + \int_0^1 dve^{-x/v}e_{v+}\phi_v(-\mu)A^-(v). \quad (27b)$$

On adding and subtracting, there results

$$\psi(x,-\mu) \pm \psi(\Delta-x,\mu) = \tau_0^\pm(x,\mu)b_0^\pm + \int_0^1 dv\tau^\pm(x,v,\mu)B^\pm(v) \quad (28a)^\pm$$

with

$$\tau_0^\pm(x,\mu) \equiv \phi_{0+}(-\mu)e^{-x/v_0} \pm e_{0-}\phi_{0+}(\mu)e^{x/v_0}$$
$$\tau^\pm(x,v,\mu) \equiv e^{-x/v}\phi_v(-\mu) \pm e^{x/v}e_{v-}\phi_v(\mu). \quad (28b,c)^\pm$$

Further, to simplify, the integral in Eq(28a)$^\pm$, we consider it as two integrals

$$I^\pm(\mu) \equiv \int_0^1 dv\tau^\pm(x,v,\mu)B^\pm(v) = I_1^\pm(\mu) + I_2^\pm(\mu) \quad (29a)^\pm$$

with

$$I_1^\pm(\mu) \equiv \int_0^1 dv\phi_v(-\mu)e^{-x/v}B^\pm(v) = \frac{c}{2}\int_0^1 dv\frac{v}{v+\mu}e^{-x/v}B^\pm(v) \quad (29b)^\pm$$

and

$$I_2^\pm(\mu) \equiv \int_0^1 dv e^{-(\Delta-x)/v}\left[\frac{cv}{2}\frac{1}{v-\mu}+\lambda(v)\delta(v-\mu)\right]B^\pm(v). \quad (29c)^\pm$$

Removing the singularity and integrating over the delta function in $(29c)^\pm$ gives

$$I_2^\pm(\mu) \equiv \frac{c}{2}\int_0^1 dv v\left[\frac{e^{-(\Delta-x)/v}-e^{-(\Delta-x)/\mu}}{v-\mu}\right]B^\pm(v)+e^{-(\Delta-x)/\mu}\frac{c}{2}\int_0^1 dv\frac{v}{v-\mu}B^\pm(v)+ \quad (30)^\pm$$
$$+\lambda(v)e^{-(\Delta-x)/\mu}B^\pm(v).$$

As above for the exiting distributions, we replace the principal value integral by $\text{Eq}(23b)^\pm$

$$\frac{c}{2}\int_0^1 dv\frac{v}{v-\mu}B^\pm(v) = -\pm\frac{c}{2}\int_0^1 dv\frac{v}{v+\mu}e_{v-}B^\pm(v)+ \quad (31a)^\pm$$
$$+\psi(\mu)-\lambda(\mu)B^\pm(\mu)-u_0^\pm(\mu)b_0^\pm,$$

and therefore

$$I_2^\pm(\mu) \equiv \frac{c}{2}\int_0^1 dv v\left[\frac{e^{-(\Delta-x)/v}-e^{-(\Delta-x)/\mu}}{v-\mu}\right]B^\pm(v)- \quad (31b)^\pm$$
$$-\pm e^{-(\Delta-x)/\mu}\frac{c}{2}\int_0^1 dv\frac{v}{v+\mu}e_{v-}B^\pm(v)+\psi(\mu)e^{-(\Delta-x)/\mu}-u_0^\pm(\mu)e^{-(\Delta-x)/\mu}b_0^\pm.$$

Introducing the two integrals into $\text{Eq}(29a)^\pm$ gives

$$I^\pm(\mu) = \frac{c}{2}\int_0^1 dv\frac{v}{v+\mu}\left[e^{-x/v}-e^{-(\Delta-x)/\mu}e^{-\Delta/v}\right]B^\pm(v)+$$
$$+\pm\frac{c}{2}\int_0^1 dv v\left[\frac{e^{-(\Delta-x)/v}-e^{-(\Delta-x)/\mu}}{v-\mu}\right]B^\pm(v)+ \quad (32)^\pm$$
$$+\pm\psi(\mu)e^{-(\Delta-x)/\mu}-\pm u_0^\pm(\mu)e^{-(\Delta-x)/\mu}b_0^\pm.$$

Thus, for $\text{Eq}(28a)^\pm$

$$T^{\pm}(x,\mu) \equiv \psi(x,-\mu) \pm \psi(\Delta-x,\mu)$$
$$= \pm\psi(\mu)e^{-(\Delta-x)/\mu} + \left[\tau_0^{\pm}(x,\mu) - \pm u_0^{\pm}e^{-(\Delta-x)/\mu}\right]b_0^{\pm} +$$
$$+ \frac{c}{2}\int_0^1 dv \frac{v}{v+\mu} e^{-x/v}\left[1 - e^{-(\Delta-x)/\mu}e^{-(\Delta-x)/v}\right]B^{\pm}(v) \qquad (33)^{\pm}$$
$$\pm \frac{c}{2}\int_0^1 dv\, v \left[\frac{e^{-(\Delta-x)/v} - e^{-(\Delta-x)/\mu}}{v-\mu}\right] B^{\pm}(v)$$

Note that the first term on the second line of Eq(33)$^{\pm}$ is the uncollided flux. Finally, adding and subtracting Eq(33)$^{\pm}$ and dividing by 2 gives the angular fluxes at $x$ and $\Delta-x$ in all directions

$$\psi(x,-\mu) = \frac{1}{2}\left[T^+(x,\mu) + T^-(x,\mu)\right]$$
$$\psi(\Delta-x,\mu) = \frac{1}{2}\left[T^+(x,\mu) - T^-(x,\mu)\right]. \qquad (34a)$$

or

$$\psi(x,-\mu) = \frac{1}{2}\left\{ \begin{array}{l} \left[\tau_0^+(x,\mu)b_0^+ + \tau_0^-(x,\mu)b_0^- - \left(u_0^+b_0^+ - u_0^-b_0^-\right)e^{-(\Delta-x)/\mu}\right] + \\ + \frac{c}{2}\int_0^1 dv \frac{v}{v+\mu}e^{-x/v}\left[1 - e^{-(\Delta-x)/\mu}e^{-(\Delta-x)/v}\right]\left[B^+(v) + B^-(v)\right] + \\ + \frac{c}{2}\int_0^1 dv\, v\left[\frac{e^{-(\Delta-x)/v} - e^{-(\Delta-x)/\mu}}{v-\mu}\right]\left[B^+(v) - B^-(v)\right] \end{array} \right\} \qquad (34b)$$

$$\psi(\Delta-x,\mu) = \frac{1}{2}\left\{ \begin{array}{l} \left[\tau_0^+(x,\mu)b_0^+ - \tau_0^-(x,\mu)b_0^- + \left(u_0^+b_0^+ - u_0^-b_0^-\right)e^{-(\Delta-x)/\mu}\right] + \\ + \frac{c}{2}\int_0^1 dv \frac{v}{v+\mu}e^{-x/v}\left[1 - e^{-(\Delta-x)/\mu}e^{-(\Delta-x)/v}\right]\left[B^+(v) - B^-(v)\right] + \\ + \frac{c}{2}\int_0^1 dv\, v\left[\frac{e^{-(\Delta-x)/v} - e^{-(\Delta-x)/\mu}}{v-\mu}\right]\left[B^+(v) + B^-(v)\right] \end{array} \right\} \qquad (34c)$$

## III. NUMERICAL IMPLEMENTATION

Numerical implementation first requires $\nu_0$ from a high precision Newton-Raphson solver applied to Eq(3c). Apparently, high precision for the transport eigenvalue $\nu_0$ is crucial to the numerical results that follow. One can show that for $c$ near and below 0.1, any purely numerical method to determine $\nu_0$ will lose significance without extended precision (beyond Quadruple Precision(QP)) including the present method. We consider neither extended precision nor computer algebra in this work in order to maintain reader accessibility to the numerical methods presented. For this reason, only $c \geq 0.1$ will be considered and we otherwise defer to the response matrix, DPN, doubling and Riccati. Next, the zeros of the modified Legendre polynomials come from expressing the recurrence for Legendre polynomials in matrix form and determining the matrix eigenvalues (zeros) when $P_N$ is set to zero. The Legendre functions are determined from recurrence or infinite series according to the stability of the recurrence relation for $Q_N$. The Legendre polynomials come from their stable recurrence relation. Finally, matrix inversions are computed by LU decomposition.

If you have followed my publications and presentations over the past 20 years, you may find it odd that I have not used Wynn-epsilon (W-e) acceleration [13] to further accelerate precision in quadrature order. The reason is that a high quadrature order is already required for LNM as is and there is little advantage of W-e acceleration.

To demonstrate the extreme precision (greater than nine digits) that the LNM can achieve, we compare results with the RM/DOM method (9) for several sample cases.

### III.A. For an Isotropic Source
The first comparison is for an isotopically distributed source on the near surface

$$\psi(0,\mu) = \psi(\mu) \equiv 1. \tag{35}$$

Each of the following tables display the benchmark from RM/DOM and include the discrepant digits underlined in bold as calculated by LNM.

Table 1a shows a high precision demonstration of the exiting flux variation [$\mu$ negative for $x = 0$ and $\mu$ positive for $x = \Delta$] with slab thickness $\Delta$ and $c = 0.9$. There are only three missed entries by one unit in the ninth place over the entire range of thicknesses considered. The quadrature order for the LNM is 600 and for RM/DOM about 300.

**Table 1a**. Exiting Flux variation with $\Delta$ for $c = 0.9$ for an isotropic source

| $\mu\backslash\Delta$ | 0.5 | 1 | 2 | 4 | 8 | 16 |
|---|---|---|---|---|---|---|
| -1.0000E+00 | 1.690394663E-01 | 2.674103351E-01 | 3.616488728E-01 | 4.081422059E-01 | 4.148431584E-01 | 4.149474557E-01 |
| -8.0000E-01 | 2.011521450E-01 | 3.078331616E-01 | 4.008507470E-01 | 4.419247759E-01 | 4.473419639E-01 | 4.474230583E-01 |
| -6.0000E-01 | 2.477252111E-01 | 3.609294436E-01 | 4.474820691E-01 | 4.815092434E-01 | 4.85787683**5**E-01 | 4.858515235E-01 |
| -4.0000E-01 | 3.200099631E-01 | 4.311708733E-01 | 5.027117752E-01 | 5.290560312E-01 | 5.324081109E-01 | 5.324584269E-01 |
| -2.0000E-01 | 4.373716428E-01 | 5.194020596E-01 | 5.691959491E-01 | 5.889949693E-01 | 5.915740025E-01 | 5.916127886E-01 |
| 0.0000E+00 | 5.916234075E-01 | 6.353636392E-01 | 6.683187157E-01 | 6.819595126E-01 | 6.837453519E-01 | 6.837722280E-01 |
| 0.0000E+00 | 2.584017054E-01 | 1.815479981E-01 | 1.000623523E-01 | 3.390798729E-02 | 4.123403075E-03 | 6.16065802**0**E-05 |
| 2.0000E-01 | 4.107668192E-01 | 2.675980096E-01 | 1.449056449E-01 | 4.895729199E-02 | 5.950585944E-03 | 8.890332236E-05 |
| 4.0000E-01 | 5.612862521E-01 | 3.664802865E-01 | 1.907915573E-01 | 6.359067276E-02 | 7.719367729E-03 | 1.153227551E-04 |
| 6.0000E-01 | 6.580998500E-01 | 4.589432518E-01 | 2.452521161E-01 | 8.100865861E-02 | 9.79344583**5**E-03 | 1.462814570E-04 |
| 8.0000E-01 | 7.214073554E-01 | 5.332715234E-01 | 3.022759024E-01 | 1.027115598E-01 | 1.243873978E-02 | 1.857172384E-04 |
| 1.0000E+00 | 7.653830029E-01 | 5.916250896E-01 | 3.565007218E-01 | 1.285236970E-01 | 1.601958683E-02 | 2.401558921E-04 |

Thus, LNM gives nearly 9-place precision in comparison to the response matrix method. The time of computation for LNM is 17s and about 5s for RM/DOM on a 2.6MHz Dell Precision PC.

Table 1b gives a demonstration but for the variation $c$ for a 1$mfp$ slab. Here, one observes the degradation of the LNM because of its inability to fully capture $v_0$ as $c$ nears 0.1. For $c = 0.1$, only 6-place precision is achievable for $N = 2500$ and 8-place precision for $N = 3000$, which is unreasonable for a 1D transport extreme benchmark. The remainder of the table required quadrature order 600 only missing four entries by one unit in the ninth place.

**Table 1b**. Exiting Flux variation with $c$ for $\Delta = 1$ for an isotropic source

| $\mu\backslash c$ | 0.1 | 0.3 | 0.5 | 0.7 | 0.9 | 0.99 |
|---|---|---|---|---|---|---|
| -1.0000E+00 | 1.50400547_2_E-02 | 5.124571442E-02 | 9.911918554E-02 | 1.660168888E-01 | 2.674103351E-01 | 3.329218166E-01 |
| -8.0000E-01 | 1.7518830_8_3E-02 | 5.955471562E-02 | 1.148847489E-01 | 1.918241465E-01 | 3.078331616E-01 | 3.825130896E-01 |
| -6.0000E-01 | 2.0928938_1_0E-02 | 7.088964428E-02 | 1.361760310E-01 | 2.262501843E-01 | 3.609294436E-01 | 4.471117240E-01 |
| -4.0000E-01 | 2.58611594_7_E-02 | 8.702979248E-02 | 1.659221624E-01 | 2.732166375E-01 | 4.311708733E-01 | 5.311207558E-01 |
| -2.0000E-01 | 3.35361494_7_E-02 | 1.112934854E-01 | 2.087387645E-01 | 3.370749686E-01 | 5.194020596E-01 | 6.317828084E-01 |
| 0.0000E+00 | 5.12827307_1_E-02 | 1.628491538E-01 | 2.904737980E-01 | 4.423922290E-01 | 6.353636392E-01 | 7.447453683E-01 |
| 0.0000E+00 | 8.0285146_2_7E-03 | 2.866250776E-02 | 5.85442675_0_E-02 | 1.045295474E-01 | 1.815479981E-01 | 2.34850861_3_E-01 |
| 2.0000E-01 | 1.8137819_3_E-02 | 4.764831699E-02 | 9.062607009E-02 | 1.56868720_8_E-01 | 2.675980096E-01 | 3.440071157E-01 |
| 4.0000E-01 | 9.5500161_2_0E-02 | 1.294549833E-01 | 1.774911944E-01 | 2.494526695E-01 | 3.664802865E-01 | 4.458002699E-01 |
| 6.0000E-01 | 2.021168_0_62E-01 | 2.352555191E-01 | 2.814495411E-01 | 3.496391717E-01 | 4.589432518E-01 | 5.323251387E-01 |
| 8.0000E-01 | 2.988610467E-01 | 3.295973198E-01 | 3.720980262E-01 | 4.343264153E-01 | 5.332715234E-01 | 5.993412930E-01 |
| 1.0000E+00 | 3.79229480_7_E-01 | 4.073575782E-01 | 4.460583899E-01 | 5.024362348E-01 | 5.916250896E-01 | 6.509775491E-01 |

Table 2 shows the flux for several interior points for a 10$mfp$ slab with $c = 0.9$ from Eqs(33)$^{\pm}$ and (34b,c). Again, in comparison with RM/DOM all but two entries agree to all nine places for a quadrature order of 600.

**Table 2**. Interior flux for $c = 0.9$ and $\Delta = 10$

| $\mu\backslash x$ | 0 | $\Delta/4$ | $\Delta/2$ | $3\Delta/4$ | $\Delta$ |
|---|---|---|---|---|---|
| -1.000E+00 | 4.149346930E-01 | 1.059596412E-01 | 2.819531781E-02 | 6.906277169E-03 | 0.000000000E+00 |
| -8.000E-01 | 4.474131699E-01 | 1.138373499E-01 | 3.033083717E-02 | 7.552989117E-03 | 0.000000000E+00 |
| -6.000E-01 | 4.858437364E-01 | 1.229773118E-01 | 3.279557816E-02 | 8.29640409_0_E-03 | 0.000000000E+00 |
| -4.000E-01 | 5.324522883E-01 | 1.337144261E-01 | 3.567748073E-02 | 9.14508935_5_E-03 | 0.000000000E+00 |
| -2.000E-01 | 5.916080563E-01 | 1.465148112E-01 | 3.909985304E-02 | 1.011916951E-02 | 0.000000000E+00 |
| 0.000E+00 | 6.837689487E-01 | 1.620521264E-01 | 4.323691369E-02 | 1.127082001E-02 | 0.000000000E+00 |
| 0.000E+00 | 1.000000000E+00 | 1.620521264E-01 | 4.323691369E-02 | 1.127082001E-02 | 1.441451176E-03 |
| 2.000E-01 | 1.000000000E+00 | 1.813652202E-01 | 4.834433965E-02 | 1.267387255E-02 | 2.080149120E-03 |
| 4.000E-01 | 1.000000000E+00 | 2.066523350E-01 | 5.481924043E-02 | 1.443494438E-02 | 2.698349736E-03 |
| 6.000E-01 | 1.000000000E+00 | 2.406803291E-01 | 6.334828112E-02 | 1.672533934E-02 | 3.422877684E-03 |
| 8.000E-01 | 1.000000000E+00 | 2.810438211E-01 | 7.489769056E-02 | 1.983815164E-02 | 4.346285746E-03 |
| 1.000E+00 | 1.000000000E+00 | 3.233270541E-01 | 9.000257612E-02 | 2.419587784E-02 | 5.612394529E-03 |

### III.B. For an Exponential Source
Now consider the normalized exponentially distributed source

$$\psi(0,\mu) = \psi(\mu) \equiv \frac{e^{-\mu}}{1-e^{-1}}. \tag{36}$$

For this case, we have no standard of comparison other than comparison to increasing quadrature and anticipating convergence.

Table 3 shows that the exponential source requires more effort than the isotropic source to achieve only 6-digit precision, which is certainly adequate for a benchmark, but not an extreme benchmark.

Table 3. For exponential Source: $\Delta = 1$, $c=0.99$

| $\mu$\N | 600 | 800 | 1000 | 1200 | 1400 | 1600 | 2000 |
|---|---|---|---|---|---|---|---|
| -1.0000E+00 | 3.324640E-01 | 3.324642E-01 | 3.324644E-01 | 3.324644E-01 | 3.324645E-01 | 3.324645E-01 | 3.324645E-01 |
| -8.0000E-01 | 3.823575E-01 | 3.823578E-01 | 3.823579E-01 | 3.823580E-01 | 3.823581E-01 | 3.823581E-01 | 3.823581E-01 |
| -6.0000E-01 | 4.476441E-01 | 4.476445E-01 | 4.476447E-01 | 4.476448E-01 | 4.476449E-01 | 4.476449E-01 | 4.476449E-01 |
| -4.0000E-01 | 5.333890E-01 | 5.333895E-01 | 5.333897E-01 | 5.333898E-01 | 5.333899E-01 | 5.333900E-01 | 5.333900E-01 |
| -2.0000E-01 | 6.394156E-01 | 6.394162E-01 | 6.394165E-01 | 6.394167E-01 | 6.394168E-01 | 6.394169E-01 | 6.394169E-01 |
| 0.0000E+00 | 7.871976E-01 | 7.871987E-01 | 7.871992E-01 | 7.871994E-01 | 7.871996E-01 | 7.871997E-01 | 7.87199**8**E-01 |
| 0.0000E+00 | 2.306036E-01 | 2.306036E-01 | 2.306036E-01 | 2.306036E-01 | 2.306037E-01 | 2.306037E-01 | 2.306037E-01 |
| 2.0000E-01 | 3.383456E-01 | 3.383457E-01 | 3.383457E-01 | 3.383458E-01 | 3.383458E-01 | 3.383458E-01 | 3.383458E-01 |
| 4.0000E-01 | 4.410439E-01 | 4.410441E-01 | 4.410442E-01 | 4.410442E-01 | 4.410442E-01 | 4.410442E-01 | 4.41044**3**E-01 |
| 6.0000E-01 | 5.285819E-01 | 5.285821E-01 | 5.285822E-01 | 5.285823E-01 | 5.285823E-01 | 5.285823E-01 | 5.285823E-01 |
| 8.0000E-01 | 5.963301E-01 | 5.963303E-01 | 5.963304E-01 | 5.963304E-01 | 5.963305E-01 | 5.963305E-01 | 5.963305E-01 |
| 1.0000E+00 | 6.484839E-01 | 6.484841E-01 | 6.484842E-01 | 6.484842E-01 | 6.484843E-01 | 6.484843E-01 | 6.484843E-01 |

### III.C For the perpendicular beam source
### III.C.1 Analytical approximation of delta function

The beam source

$$\psi(0,\mu) = \psi(\mu) \equiv \delta(v - \mu_0) \quad (37)$$

is the most challenging for the LNM. The primary difficulty is the numerical representation of the delta function on the RHS of Eq(8a)$^\pm$. Theoretically, the LNM solves SIEs to construct the flux from the expansion coefficients but not directly for the exiting flux. So we start at a more basic construction rather than at the solution itself, which is most likely why we have difficulties with the delta function since LNM does not require integration over the delta function. Ideally, the delta function should be carried along analytically in any theoretical manipulation as will be shown; but first, we consider an obvious analytical approximation of the delta function.

Thus, we adopt the following formal/analytical representation:

$$\delta(v - \mu_0) = \sum_{l=0}^{\infty} \frac{2l+1}{2} P_l(v) P_l(\mu_0), \quad (38a)$$

which leads to

$$\delta(v - \mu_0) = 2\delta((2v-1) - (2\mu_0 - 1)) = \sum_{l=0}^{\infty} (2l+1) P_l^*(v) P_l^*(\mu_0). \quad (38b)$$

If we truncate at $N$

$$\delta(v - \mu_0) \simeq \sum_{l=0}^{N} (2l+1) P_l^*(v) P_l^*(\mu_0), \quad (39a)$$

then from the Darboux formula[14]

$$\delta(v - \mu_0) \simeq (N+1)\left[\frac{P^*_{N+1}(v) P^*_N(\mu_0) - P^*_N(v) P^*_{N+1}(\mu_0)}{v - \mu_0}\right] \quad (39b)$$

and for $v = v_j$ and if $\mu_0 = 1$, since $P^*_N(v_j) = 0$ and $P^*_N(1) = 1$

$$\delta(v_j - 1) \simeq (N+1)\left[\frac{P^*_{N+1}(v_j)}{v_j - 1}\right]. \quad (39c)$$

Furthermore since

$$P^*_{N+1}(v_j) = \frac{1}{(N+1)Q^*_N(v_j)} \quad (40a)$$

Eq(39c) becomes

$$\delta(v_j - 1) \simeq \left[\frac{1}{Q^*_{N+1}(v_j)(v_j - 1)}\right]. \quad (40b)$$

**III.C.2 Theoretical substitution**
Starting with Eq(8a)$^\pm$ for $i = 1$,

$$\lambda(v) B_1^\pm(v) + \frac{c}{2}\int_0^1 dv' \frac{v'}{v' - v} B_1^\pm(v') \pm \frac{c}{2}\int_0^1 dv' \frac{v'}{v' + v} e_{v'_-} B_1^\pm(v') = \delta(v - \mu_0), \quad (41a)$$

one can eliminate the delta function by the substitution

$$B_1^\pm(v) = \frac{1}{\lambda(v)}\delta(v - \mu_0) + C_1^\pm(v). \quad (41b)$$

Eq(41a) becomes

$$\lambda(v)C_1^\pm(v) + \frac{c}{2}\int_0^1 dv' \frac{v'}{v'-v} C_1^\pm(v') \pm \frac{c}{2}\int_0^1 dv' \frac{v'}{v'+v} e_{v'-} C_1^\pm(v') =$$
$$= -\frac{c}{2}\frac{\mu_0}{\lambda(\mu_0)}\left[\frac{1}{\mu_0 - v} \pm \frac{e_{\mu_0}}{\mu_0 + v}\right]. \tag{42a}$$

For the perpendicular beam ($\mu_0 = 1$), since $\lambda(\mu_0) \to \infty$, Eq(42) gives the solution

$$C_1^\pm(v) = 0, \quad 0 \leq v \leq 1, \tag{42b}$$

and therefore from Eq(41b)

$$B_1^\pm(v) = 0. \tag{42c}$$

The exiting fluxes for a perpendicular are from Eqs(25)

$$\psi(0,-\mu) = \frac{1}{2}\left\{\begin{aligned}&\left[\tau_0^+(\mu) - u_0^+ e_{\mu-}\right]b_0^+ + \left[\tau_0^-(\mu) + u_0^- e_{\mu-}\right]b_0^- +\\ &-\frac{c}{2}\int_0^1 dv \frac{v}{v+\mu}\left[1 - e_{\mu-}e_{v-}\right]\left[b_0^+ B_2^+(v) + b_0^- B_2^-(v)\right] +\\ &-\frac{c}{2}\int_0^1 dvv\left[\frac{e_{v-} - e_{\mu-}}{v-\mu}\right]\left[b_0^+ B_2^+(v) - b_0^- B_2^-(v)\right]\end{aligned}\right\} \tag{43a}$$

$$\psi(\Delta,\mu) = \delta(\mu-\mu_0)e_{\mu_0-} +$$
$$+\frac{1}{2}\left\{\begin{aligned}&\left[\tau_0^+(\mu) - u_0^+ e_{\mu-}\right]b_0^+ - \left[\tau_0^-(\mu) + u_0^- e_{\mu-}\right]b_0^- +\\ &-\frac{c}{2}\int_0^1 dv \frac{v}{v+\mu}\left[1 - e_{\mu-}e_{v-}\right]\left[b_0^+ B_2^+(v) - b_0^- B_2^-(v)\right] +\\ &-\frac{c}{2}\int_0^1 dvv\left[\frac{e_{v-} - e_{\mu-}}{v-\mu}\right]\left[b_0^+ B_2^+(v) + b_0^- B_2^-(v)\right]\end{aligned}\right\}, \tag{43b}$$

and from Eq(19)$^\pm$

$$b_0^\pm = \left[ N_{0+} + T_1^\pm - \sum_{j=1}^{N} \omega_j T_{2j}^\pm B_{2j} \right]^{-1} J_0^+. \qquad (43c)^\pm$$

In Table 4, we have LNM results from Eq(40b) for increasing $N$ showing convergence. The highlighted digits are in disagreement with RM/DOM. We observe that at best, we get five- place precision confirming the difficulty of the delta function source. Five-place precision however, is adequate for most benchmarking applications.

**Table 4.** Exiting Flux variation with $\Delta$ for $c = 0.9$ for a perpendicular beam source

| $\mu$\\$N$ | 500 | 1000 | 1500 | 2000 |
|---|---|---|---|---|
| -1.0000E+00 | 4.200004E-01 | 4.200013E-01 | 4.200014E-01 | 4.20001<u>5</u>E-01 |
| -8.0000E-01 | 4.789422E-01 | 4.789433E-01 | 4.789435E-01 | 4.78943<u>6</u>E-01 |
| -6.0000E-01 | 5.530881E-01 | 5.530894E-01 | 5.530896E-01 | 5.53089<u>7</u>E-01 |
| -4.0000E-01 | 6.425319E-01 | 6.425333E-01 | 6.425336E-01 | 6.42533<u>7</u>E-01 |
| -2.0000E-01 | 7.266104E-01 | 7.266121E-01 | 7.266124E-01 | 7.26612<u>6</u>E-01 |
| 0.0000E+00 | 7.187397E-01 | 7.187415E-01 | 7.187419E-01 | 7.18742<u>0</u>E-01 |
| 0.0000E+00 | 3.722744E-01 | 3.722750E-01 | 3.722751E-01 | 3.722752E-01 |
| 2.0000E-01 | 4.966540E-01 | 4.966549E-01 | 4.966551E-01 | 4.966552E-01 |
| 4.0000E-01 | 5.104030E-01 | 5.104040E-01 | 5.104042E-01 | 5.104043E-01 |
| 6.0000E-01 | 4.710710E-01 | 4.710720E-01 | 4.710721E-01 | 4.71072<u>2</u>E-01 |
| 8.0000E-01 | 4.237652E-01 | 4.237661E-01 | 4.237662E-01 | 4.23766<u>3</u>E-01 |
| 1.0000E+00 | 3.805283E-01 | 3.805291E-01 | 3.805292E-01 | 3.805293E-01 |

### IV. CONCLUSION

A new numerical neutron transport solution, called the Lagrange order $N$ Method (LNM), for a 1D slab with isotropic scattering is established for a surface source. The solution features Lagrange interpolation in combination with Gauss Legendre quadrature. The method seems stable, at least for an isotropic source giving what is believed to be the first ever 1D slab benchmark to nine places. The LNM approach is quite revolutionary as it relatively simply evaluates Case's elegant solution without applying other than full-range orthogonality. A remaining issue is how to more effectively treat the beam source. One can approximate the delta function as its formal Legendre expansion, but only five digits of precision result. Thus, one cannot consider this a true extreme benchmark until the beam source is resolved, which will require an alternative procedure for the expansion coefficients (submitted[16]). However, there is no escaping that resolution of a benchmark for $c < 0.1$ will require multi-precision arithmetic.

I leave you with a thought expressed by Professor Norman McCormick, a noted transport theorist who is an expert in Caseology, regarding SEEs [15]

> *…., it seems safe to state that for someone interested only in solving practical nuclear engineering problems, eigenfunction expansions will have little appeal. If, however, one also seeks a thorough mathematical understanding of those problems, such a study may be quite rewarding and may provide excellent training in applied mathematics.*